\documentclass[9pt,twocolumn,twoside]{osajnl}

\journal{ol} 

\setboolean{shortarticle}{true}

\title{Passive Optical Phase Noise Cancellation}

\author[*]{Liang Hu}
\author[ ]{Xueyang Tian}
\author[ ]{Guiling Wu}
\author[ ]{Jianping Chen}

\affil[ ]{State Key Laboratory of Advanced Optical Communication Systems and Networks, Department of Electronic Engineering, Shanghai Jiao Tong University, Shanghai 200240, China}

\affil[*]{Corresponding author: liang.hu@sjtu.edu.cn}

\begin{abstract}
We report on the realization of an optical phase noise cancellation technique by passively embedding the optical phase information into a radio frequency (RF) signal and shifting the optical frequency with the amount of phase noise introduced by optical phase perturbations. Neither phase discrimination nor active phase tracking is required due to the open-loop design, mitigating some technical problems, such as the limited compensation speed and the finite phase/timing jitter, in conventional phase noise cancellation.  We experimentally demonstrate that this technique maintains the same delay-limited  bandwidth and phase noise suppression capability as in conventional techniques, but significantly shortens the response speed and phase recovery time.  Passive decoupling optical phase perturbation represents a powerful technique in the domains of optical frequency standard comparisons and clockworks for future optical atomic clocks, which are now under serious investigation for a potential redefinition of the International Time Scale.
\end{abstract}

\setboolean{displaycopyright}{true}

\begin{document}

\maketitle

Clocks ticking at optical frequencies slice time into much finer intervals than do microwave clocks and thus provide unprecedented accuracy, surpassing that of previously available clock systems by more than a few orders of magnitude \cite{ludlow2015optical, mcgrew2018atomic, marti2018imaging}.  Precise clock networks have the potential to enable dramatic improvements in precision navigation and timing, radio astronomy \cite{clivati2017vlbi, wang2015square}, clock-based geodesy \cite{grotti2018geodesy}, testing of fundamental physics \cite{lisdat2016clock}, geological disaster observation \cite{marra2018ultrastable} and even future searching for dark matter and gravitational-wave detection \cite{kolkowitz2016gravitational, hu2017atom}.  At the same time, precision comparisons of different clocks will also be necessary for a future redefinition of the second in the International System of Units (SI)  \cite{grebing2016realization, mcgrew2019towards}. The current well-established, satellite-based techniques for microwave dissemination are not adequate for comparing optical clocks. Taking advantages of the low loss, high stability, large bandwidth, and immunity to electromagnetic interference, the use of optical fiber, today's best optical frequency transfer technique, enables the comparison of state-of-the-art optical atomic clocks and achievement of  coherent optical phase transfer over large distances at will, up to thousands of kilometers, preserving their coherence at an extremely high level \cite{PhysRevLett.111.110801, Calonico2014, deng2016coherent}. 

\begin{figure*}[htbp]
\centering
\includegraphics[width=0.95\linewidth]{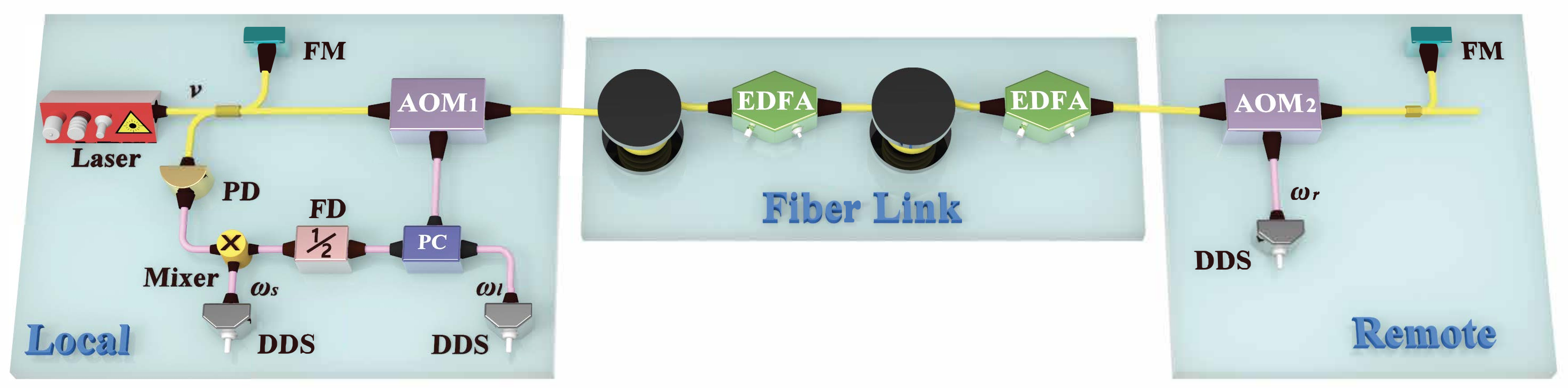}
\caption{Schematic diagram of our optical frequency dissemination using a technique with passive phase noise cancellation in an optical fiber. AOM: acousto-optic modulator, FM: Faraday mirror, DDS: direct-digital synthesizer, EDFA: erbium-doped-fiber-amplifier, PD: photo-detector, FD: frequency divider, PC: power combiner}
\label{fig1}
\end{figure*}

Optical phase noise in the transmitted frequency introduced by thermal and acoustic fluctuations in the fiber link can be cancelled by either physically modulating the path length or by intentionally shifting the frequency. In practical implementations of these methods, the phase variations caused by the changes in the fiber link should be extracted and used to drive tunable devices for phase variation compensation, such as piezoelectric fiber stretchers and voltage-controlled oscillators (VCOs) as demonstrated by Ma \textit{et al}. in 1994 \cite{ma1994delivering}. The main drawback related to this conventional approach is the need for complex circuits  to extract the phase error and drive the devices for phase correction in real time, e.g., frequency dividers with a high ratio needed to avoid the cycle slips. Although the method based on VCOs principally has an endless compensation range, achieving short settling time of stabilization locks, namely, a fast compensation speed, is difficult and is a great challenge even for dedicated servo amplifiers. The characteristic of the short settling time is extremely important for the applications, such as transferring the pulse light and delivery of light over free-space where the interruptions frequently occur \cite{Falke2012, giorgetta2013optical, kang2019free}.

While radio frequency (RF) transfer by multiplexing and dividing the transferred RF signal has been proposed and experimentally demonstrated \cite{pan2016passive}, these solutions are not applicable to optical frequency dissemination, in particular, for fiber-based optical frequency dissemination by multiplexing and dividing the transferred optical signal itself. In view of the upcoming redefinition of the second and future applications of optical atomic clocks, technological developments  toward more efficient, simple, and robust devices are needed.  In this Letter, we propose and demonstrate a novel approach for highly stable dissemination of the phase of an optical frequency reference over an optical-fiber link without the need for the active stabilization of the optical path length. Here neither active phase discrimination nor dynamic phase tracking/compensation is required in our open-loop design, enabling the proposed approach to be robust, compact, cost-effective, and easy to implement.

Figure \ref{fig1} depicts a schematic diagram of our optical frequency transfer using a technique with passive phase noise cancellation in an optical fiber. Part of the light with an angular frequency of $\nu$ is split off to serve as a local oscillator. The remaining part of this reference light is launched into the link via an acousto-optic modulator 1 (AOM$_1$) working at an angular frequency of $\omega_l$ (downshifted mode) and then propagated to the remote end via the optical fiber link, where it encounters AOM$_2$ (upshifted mode), which provides a fixed angular frequency shift $\omega_r$ ($\omega_r>\omega_l$) to distinguish light reflected at the remote end from stray reflections along the link \cite{ma1994delivering, williams2008high}. Most of the light is retroreflected from the remote end via a Faraday mirror, emerges from the start of the fiber, and is heterodyned against the local oscillator light to yield a $2(-\omega_l+\omega_r)$ RF signal with phase noise of $2\phi_p$ at the local photo-detector (PD). Here, we assume that phase noise $\phi_p$ in the forward and backward optical paths is equal. Thus, this RF signal includes twice the phase noise of the one-way transmitted optical frequency. To further avoid light reflected at the remote site from stray reflections generated along the fiber link,  the detected signal is mixed with an assistant RF signal $\omega_s$ ($\omega_s<-\omega_l+\omega_r$), and then,  the angular frequency of the upper sideband of the mixed signal is filtered out  and then divided by a factor of 2. Afterwards, its output together with $\omega_l$  is fed into the same AOM$_1$ with the assistance of an RF power combiner, resulting in a desirable optical signal with angular frequency $\nu+\omega_l-\omega_r-\frac{1}{2}\omega_s$ and phase $-\phi_p$ at the local site. With this configuration, the first and third pass forward optical frequencies differing by $2\omega_l-\frac{1}{2}\omega_s-\omega_r$ follow the same optical path; thus, phase fluctuations caused, for example, by acoustic  vibrations or temperature fluctuations are common mode, and do not degrade the optical frequency transfer performance. We carefully engineered the frequency map ($\omega_l$, $\omega_r$, $\omega_s$) to ensure that the signals were sufficiently far apart to be easily filtered and separated. Each filtered signal was amplified and tracked using a tracking oscillator with a typical bandwidth of $\sim$100 kHz. In this way, the remote end will automatically receive a stabilized optical frequency signal with an angular frequency of $\nu+\omega_l-\frac{1}{2}\omega_s$.

Note that the above description does not take the propagation delay of the fiber link into account. As noted by Williams \textit{et al.},  the capability of the phase noise rejection will be limited by the propagation delay \cite{williams2008high}. The residual phase noise power spectral density (PSD), $S_{\text{remote}}(f)$, at the remote end in terms of the single-pass free-running phase noise PSD, $S_{\text{fiber}}(f)$, and the propagation delay in the fiber, $\tau_0$, for our proposed optical frequency transfer  can be calculated as,
\begin{equation}
S_{\text{remote}}(f)\simeq\frac{1}{3}(2\pi f\tau_0)^2 S_{\text{fiber}}(f).
\label{eq1}
\end{equation}

To verify the principle of the proposed scheme, a 145 km optical fiber spool is applied in the laboratory frequency dissemination system.  To overcome the one-way attenuation of more than 30 dB, two low noise erbium-doped fiber amplifier (EDFA) systems are evenly distributed over the entire link length, with the amplifiers located at the middle and end of the fiber link, respectively as depicted in Fig. \ref{fig1}.  We built an interferometer by using free-space optics, in which the light can propagate in the optical fiber only when the local oscillator and transferred signal paths are colocated. In comparison with fiber-based interferometers \cite{Stefani:15}, the out-of-loop phase noise can be easily controlled by using free-space optics \cite{williams2008high}.  All optical components are housed in an aluminum box filled with foam to reduce their sensitivity to temperature and acoustic noise.  Here we set $\omega_l=45$ MHz, $\omega_r=75$ MHz and $\omega_s=10$ MHz, resulting in out-of-loop beat notes of 30 MHz and 40 MHz for the free-running link and the stabilized link, respectively. The frequency selection principle is mainly determined by the bandwidth of AOM$_1$ and RF bandpass filters. Here, the 3 dB bandwidths of AOM$_1$ and the RF bandpass filters are $\sim20$ MHz  and $\sim5$ MHz. The beat notes of the out-of-loop RF signals after electronic amplification and filtering with tracking oscillators, are simultaneously recorded by a dead-time free, high-resolution frequency counter (K$+$K FXE) operating in $\Pi$-type with a gate time of 1 s. Due to the absence of any frequency dividers with a high ratio at the local site, as adopted in conventional techniques, the rate of cycle slips is as low as a few per day under typical laboratory conditions, and the signal phase can be continuously tracked for a few days. Routine operation over several days requires regular adjustment of the polarization. As a reference, we measured the residual noise of the interferometer by replacing the fiber spool with a fixed attenuator of equivalent loss and leaving the $1$ m patch cord in place, establishing an upper bound of the noise introduced by the interferometer.


\begin{figure*}[htbp]
\centering
\includegraphics[width=0.95\linewidth]{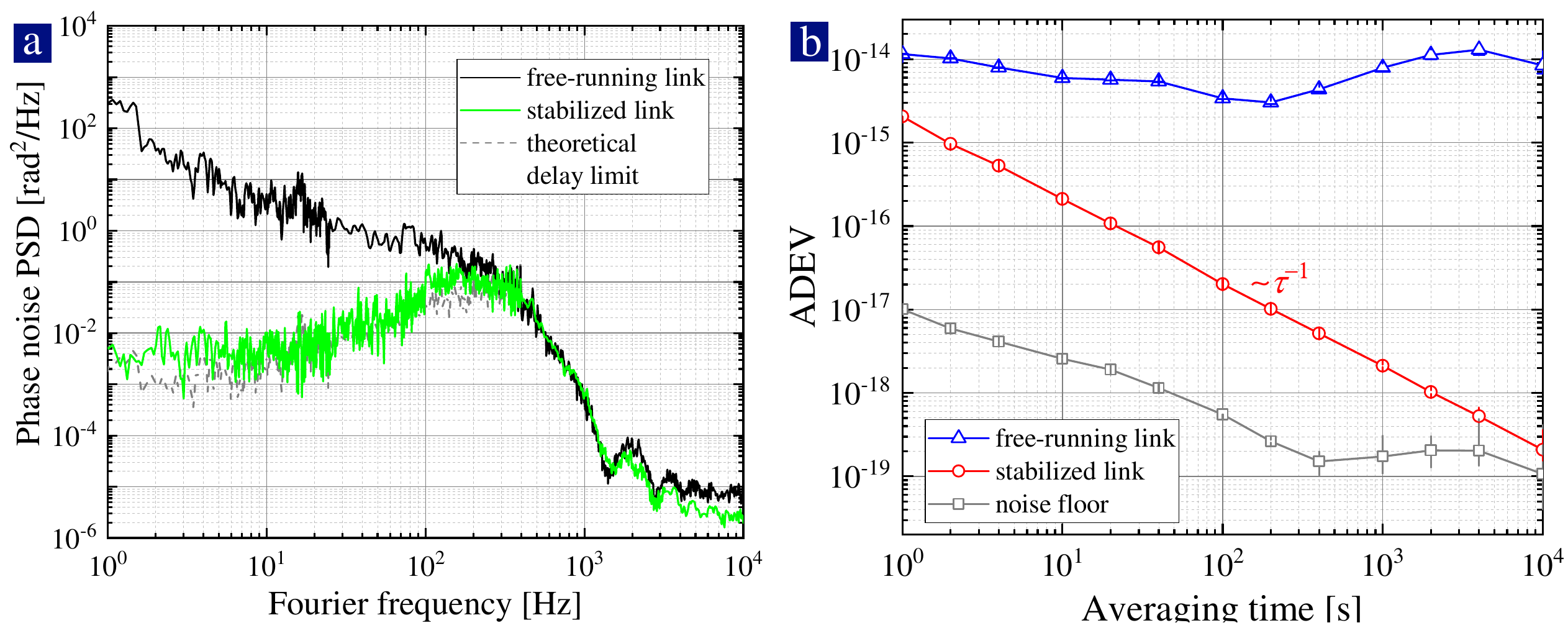}
\caption{(a) Measured phase noise PSD of the 145 km free-running fiber link (black curve) and stabilized link (green curve) with the proposed phase noise cancellation. The gray line is the theoretical prediction based on Eq. \ref{eq1}. (b) Measured fractional frequency instability of the 145 km free-running fiber link (blue triangles) and stabilized link derived from a nonaveraging ($\Pi$-type) frequency counter expressed as the ADEV (red circles). The measured noise floor of the interferometer is also shown (gray squares).}
\label{fig2}
\end{figure*}

Figure \ref{fig2}(a) shows the measured phase noise PSD for the stabilized fiber link (green curve) and the free-running link (black curve), for which no phase noise cancellation was applied.   We find that the phase noise PSD of our free-running fiber link approximately follows a power-law dependence, $S_{\phi}(f)\simeq h_{1}f^{-1}$ rad$^2$/Hz, for $f<300$ Hz, indicating that flicker phase  noise is dominant in the free-running fiber link, where $h_{1}\simeq 65$ in our fiber link. According to the theory of the delay-limited phase noise PSD illustrated in Eq. \ref{eq1}, phase noise cancellation suppresses the noise within the delay-limited bandwidth proportional to $f^2$. Consequently, the residual phase noise of the stabilized link features a $S_{\phi}(f)= h_3f^1$ rad$^2$/Hz dependency with the delay-limited bandwidth of $1/4\tau_0\simeq300$ Hz. In our experiment, $h_3\simeq6.3\times10^{-4}$. From Fig. \ref{fig2}(a), we also find the measured stabilized phase noise to be in good agreement with the theoretical prediction (gray dashed curve)  as indicated in Eq. \ref{eq1}. More importantly, strong servo bumps that appear in the conventional active phase cancellation techniques are eliminated in our phase noise cancellation scheme \cite{williams2008high, Calonico2014}. This is an important characteristic for reducing the integrated phase noise of the transferred light and preserving the transferred light coherence over a longer fiber link, as discussed in \cite{williams2008high, Calonico2014}.

\begin{figure}[htbp]
\centering
\includegraphics[width=1\linewidth]{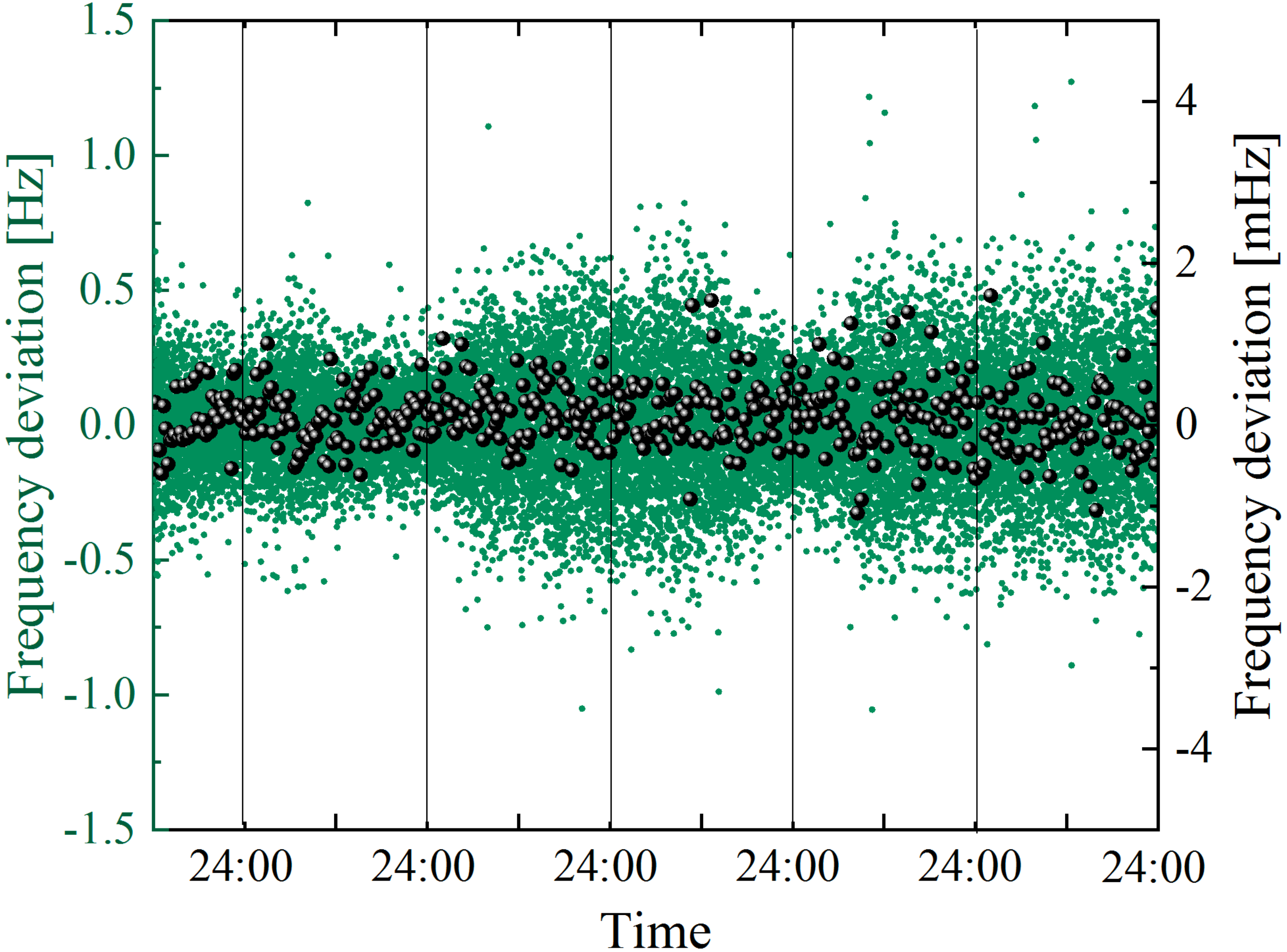}
\caption{Five-day frequency comparison between sent and transferred frequencies after 145 km. Data was taken with dead-time-free $\Pi$-type frequency counters with a gate time of 1 s (blue data points, left frequency axis). The arithmetic means of all cycle-slip free 1,000 s intervals have been computed. From the resulting 475 data points (black dots, right frequency axis, enlarged scale) a fractional difference between the sent and transferred frequencies of $-0.63\pm 2.4\times10^{-18}$ is calculated.}
\label{fig3}
\end{figure}

Complementary to the frequency-domain feature, a time-domain characterization of  the fractional frequency stability over a five-day-long measurement was performed. Figure \ref{fig2} (b) indicates the instability of the transferred frequency of the free-running link (blue triangles) as well as that of the stabilized link (red circles) expressed as overlapping Allan deviation (ADEV) recorded by $\Pi$-type frequency counter, respectively. The data from which the instability is calculated and sampled at a 1 kHz rate and a gate time of 1 s, and therefore, the equivalent measurement bandwidth is 0.5 Hz \cite{dawkins2007considerations}.  The free-running link fluctuates in the $10^{-14}$ range, with an overall mean of the fractional frequency offset of here $\sim7.5\times10^{-15}$. We obtain a value for the ADEV of $2.1\times10^{-15}$ at 1 s for the stabilized link, which scales down to $2.0\times10^{-19}$ for the integration time of 10, 000 s with a slope of $\tau^{-1}$.

Concerning the frequency accuracy, the transferred frequency could be subjected to a systematic and stable offset, for example, due to the propagation delay, which would not be revealed in the instability analysis. After measuring the transferred frequency with the $\Pi$-type frequency counter with a gate time of 1 s over a period of five days, we calculated the unweighted mean values for all cycle-slip free 1,000 s long segments. As shown in Fig. \ref{fig3}, the frequency data throughout the total measurement time is  475,419 s. Note that human activity in the laboratory causes a slight time dependence of the fiber link performance. Figure \ref{fig3} shows the difference between sent and transferred frequencies for a gate time of 1 s and the 1,000 s data subsets. The 475 data points (right axis in Fig. \ref{fig3}) have an arithmetic mean of $12.1$ $\mu$Hz ($0.63\times10^{-18}$) and a standard deviation of $85.8$ $\mu$Hz ($2.4\times10^{-18}$). Considering that the long-term stability is mainly limited by the flicker frequency noise,  as shown by the instability expressed by the ADEV shown in Fig. \ref{fig2}(b),  we can conservatively constrain the deviations from the expected frequency value to be less than $2.0\times10^{-19}$ and no frequency bias was observed within the statistical uncertainty of the data set. Based on this work, the extension to long-haul links is straightforward by inserting more amplifiers or by dividing the main link into subsections \cite{raupach2014optical, lopez2010cascaded}.

\begin{figure}[htbp]
\centering
\includegraphics[width=0.95\linewidth]{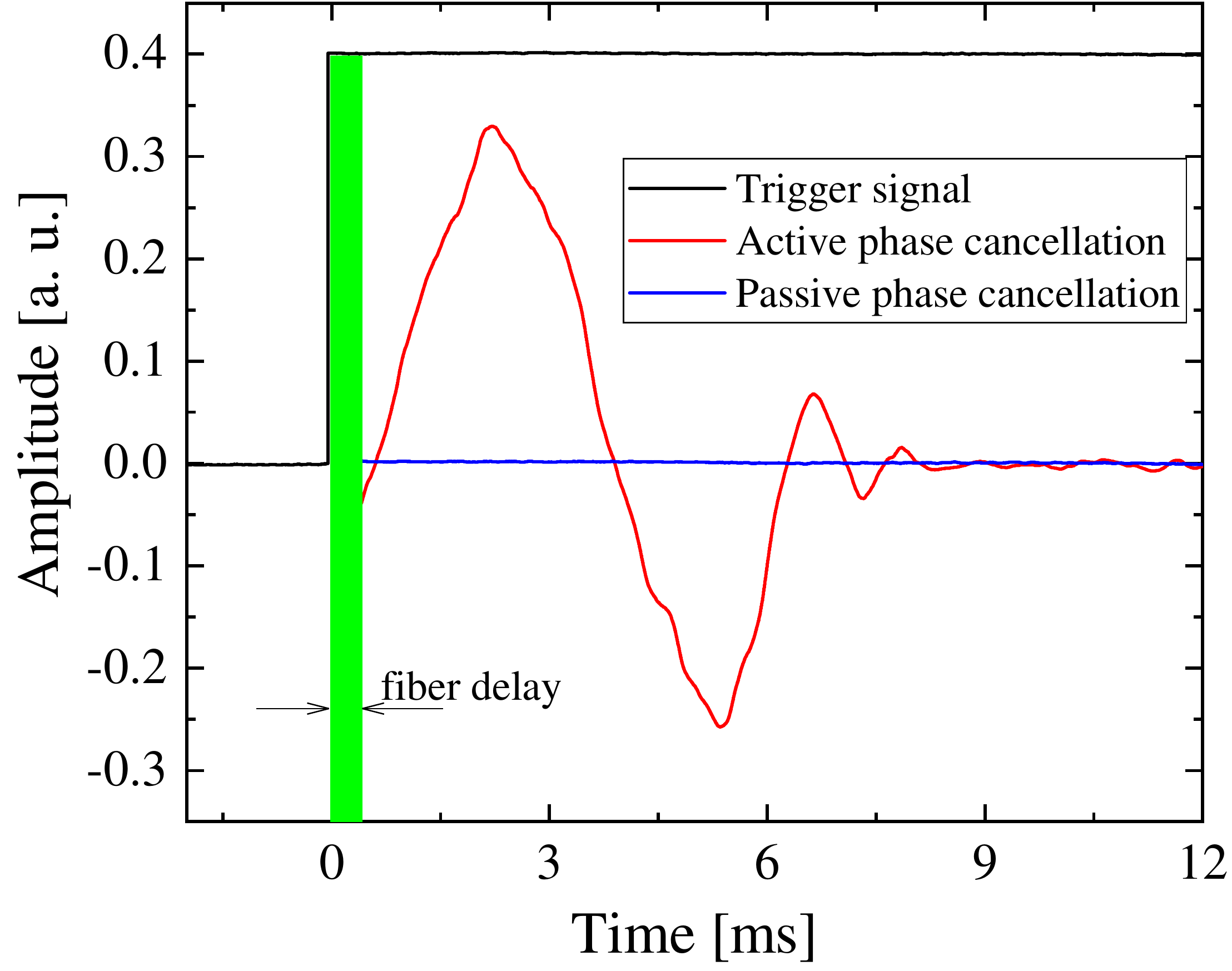}
\caption{Phase recovery time of 20 km optical path length stabilization with the proposed and conventional phase noise cancellation, respectively. A delay is introduced between the TTL signal for switching the light on at 0 ms and the activation of the phase stabilization at $4\tau_0=0.4$ ms as illustrated by the green area. The black curve indicates the TTL trigger signal. Voltage generated by mixing down the out-of-loop beat to the dc is shown for the switching-on of the active length stabilization for both the conventional (red curve) and the proposed (blue curve) phase noise correction techniques.}
\label{fig4}
\end{figure}

Unlike all other optical frequency transfer schemes with conventional phase noise correction, the scheme proposed here can significantly shorten the settling time of stabilization locks. To examine this feature, we implemented two kinds of optical frequency dissemination schemes with the proposed  and conventional phase noise cancellation over a 20 km fiber link, respectively. The configuration of optical frequency transfer with conventional phase noise correction is similar to that in \cite{PhysRevLett.111.110801, Calonico2014, deng2016coherent}, and a high-speed servo controller (Newport, LB1005-S) was used. We observed that the settling time of optical path length stabilization with conventional phase noise cancellation (red curve) is as long as $\sim8$ ms whereas the proposed implementation (blue curve) executes much faster settling time only limited by the fiber propagation delay as depicted in Fig. \ref{fig4}. This could be very beneficial for delivering pulsed light and for optical frequency transfer over free-space where the interruptions frequently occur, introduced by the atmospheric turbulence \cite{Falke2012, giorgetta2013optical, kang2019free}. Consequently, the proposed phase noise cancellation technique is significantly less sensitive to large environmental perturbations than the conventional technique.

In conclusion, we reported the first noise characterization of optical frequency dissemination over  fiber links using passively decoupling optical phase noise introduced by phase perturbations. Only frequency mixing and shifting were used, and no active mechanism was involved, enabling the fiber optical phase transfer system compact, cost-effective, and easy to implement. As we have shown, this type of optical frequency transfer would allow us to reach the same sensitivity as in the conventional optical frequency transfer techniques but with significantly reduced requirements on electronic phase noise. Moreover, we identified the mechanism with the fast response speed and phase recovery time behind the proposed phase noise cancellation technique. 


Going beyond the proof-of-principle experiment conducted under optimized laboratory conditions,  the new optical frequency transfer scheme enables a variety of applications in fundamental physics, such as tests of general relativity and searching for beyond-standard-model physics \cite{safronova2018search} as well as in optical frequency standard comparisons for future optical atomic clocks, which are now under serious consideration for a potential redefinition of the International Time Scale \cite{mcgrew2019towards, grebing2016realization}.


\medskip
\noindent\textbf{Funding.} This research was supported in part by the National Natural Science Foundation of China (NSFC) (61905143, 61627817, 61535006), and in part by science and technology project of State Grid Corporation of China (No. SGSHJX00KXJS1901531).

\medskip
\noindent\textbf{Disclosures.} The authors declare no conflicts of interest.

\bibliography{Optics}





\ifthenelse{\equal{\journalref}{aop}}{%
\section*{Author Biographies}
\begingroup
\setlength\intextsep{0pt}
\begin{minipage}[t][6.3cm][t]{1.0\textwidth} 
  \begin{wrapfigure}{L}{0.25\textwidth}
    \includegraphics[width=0.25\textwidth]{john_smith.eps}
  \end{wrapfigure}
  \noindent
  {\bfseries John Smith} received his BSc (Mathematics) in 2000 from The University of Maryland. His research interests include lasers and optics.
\end{minipage}
\begin{minipage}{1.0\textwidth}
  \begin{wrapfigure}{L}{0.25\textwidth}
    \includegraphics[width=0.25\textwidth]{alice_smith.eps}
  \end{wrapfigure}
  \noindent
  {\bfseries Alice Smith} also received her BSc (Mathematics) in 2000 from The University of Maryland. Her research interests also include lasers and optics.
\end{minipage}
\endgroup
}{}

\end{document}